\documentclass[runningheads]{llncs}

\usepackage{graphicx}

\usepackage{soulutf8, color}
\usepackage{comment}
\usepackage{booktabs}
\usepackage{multirow}
\usepackage{amsmath}

\usepackage{lineno}

\usepackage{fontawesome5}

\newcommand{\pl}[2]{#1\protect\raisebox{0.25ex}{\protect\scalebox{0.67}{\textbf{+}}}#2}
\newcommand{\mi}[2]{#1\protect\raisebox{0.2ex}{-}#2}

\begin{document}

\title{Improving AMD diagnosis by the simultaneous identification of associated retinal lesions}

\titlerunning{Improving AMD diagnosis by the identification of associated retinal lesions}

\author{
    {José Morano}\inst{1,2}\textsuperscript{(\faEnvelope[regular])}\orcidID{{0000-0003-3785-8185}}
\and
{Álvaro S. Hervella}\inst{1,2}\orcidID{{0000-0002-9080-9836}}
\and
{José Rouco}\inst{1,2}\orcidID{{0000-0003-4407-9091}}
\and
{Jorge Novo}\inst{1,2}\orcidID{{0000-0002-0125-3064}}
\and
{José I. Fernández-Vigo}\inst{3,4}\orcidID{{0000-0001-8745-3464}}
\and
{Marcos Ortega}\inst{1,2}\orcidID{{0000-0002-2798-0788}}
}

\authorrunning{{J. Morano} et al.}


\institute{
{Centro de Investigación CITIC, Universidade da Coruña, A Coruña, Spain}
\and
{VARPA Research Group, Instituto de Investigación Biomédica de A Coruña (INIBIC), Universidade da Coruña, A Coruña, Spain}\\
\email{
{\{j.morano,a.suarezh,jrouco,jnovo,mortega\}@udc.es}
}
\and
{Department of Ophthalmology, Hospital Clínico San Carlos, Instituto de Investigación Sanitaria (IdISSC), Madrid, Spain}
\and
{Department of Ophthalmology, Centro Internacional de Oftalmología Avanzada, Madrid, Spain}\\
\email{
{jfvigo@hotmail.com}
}
}

\maketitle              



\begin{abstract}
    Age-related Macular Degeneration (AMD) is the predominant cause of blindness in developed countries, specially in elderly people.
    Moreover, its prevalence is increasing due to the global population ageing.
    In this scenario, early detection is crucial to avert later vision impairment.
    Nonetheless, implementing large-scale screening programmes is usually not viable, since the population at-risk is large and the analysis must be performed by expert clinicians.
    Also, the diagnosis of AMD is considered to be particularly difficult, as it is characterized by many different lesions that, in many cases, resemble those of other macular diseases.
    To overcome these issues, several works have proposed automatic methods for the detection of AMD in retinography images, the most widely used modality for the screening of the disease.
    Nowadays, most of these works use Convolutional Neural Networks (CNNs) for the binary classification of images into AMD and non-AMD classes.
    In this work, we propose a novel approach based on CNNs that simultaneously performs AMD diagnosis and the classification of its potential lesions.
    This latter secondary task has not yet been addressed in this domain, and provides complementary useful information that improves the diagnosis performance and helps understanding the decision.
    A CNN model is trained using retinography images with image-level labels for both AMD and lesion presence, which are relatively easy to obtain.
    The experiments conducted in several public datasets show that the proposed approach improves the detection of AMD, while achieving satisfactory results in the identification of most lesions.

\end{abstract}

\section{Introduction}

Age-related Macular Degeneration (AMD) is a degenerative disorder affecting the macula. Nowadays, it is the most frequent cause of blindness in developed countries, specially in people over 60 years old~\cite{Wong_LGH_2014}. 
Worldwide, an estimated 8.7\% of blindness cases are caused by AMD~\cite{Kanski_Elsevier_2011}, and it is expected to increase due to the global population ageing.

Conventionally, the clinical classification of AMD comprises 5 classes related to its developmental stage~\cite{Kanski_Elsevier_2011}: (1) no apparent ageing changes, (2) normal ageing changes, (3) early AMD, (4) intermediate AMD and (5) late AMD. 
The late stage is generally divided into \textit{dry} and \textit{wet} AMD types, of which the first is the most common (about $90\%$ of the people diagnosed with AMD present the dry type)~\cite{Kanski_Elsevier_2011}.
All these stages are characterized by the presence of certain lesions within a distance of two optic disc diameters from the fovea of either eye.
\cite{Bird_SO_1995,Kanski_Elsevier_2011}. 
Thus, early AMD is characterized by drusen, intermediate AMD, by pigmentary abnormalities (associated to drusen), wet AMD, by choroidal neovascularization and pigment epithelial detachment, and dry AMD, by geographic atrophy.
Other less common signs of wet AMD are hemorrhages and exudates in or around the macula.
Also, it has been reported that untreated choroidal neovascularization occasionally cause a disciform scar under the macula.

Ophthalmologists can identify the presence of AMD by examining retinography (also called color fundus photography) images and optical coherence tomography (OCT) images~\cite{Ting_PRER_2019}.
OCT is more appropriate for accurate grading on diagnosed patients.
Retinography, instead, is more convenient for large-scale screening and early detection programmes for the at-risk population, due to its affordability and widespread availability.
However, implementing this type of programmes for diseases like AMD is usually not viable, since the population at risk is large and the analysis of color fundus images is challenging, as AMD is characterized by many different lesions that, in many cases, resemble those of other macular disorders~\cite{Saksens_PRER_2014}.
This forces such analysis to be performed either by clinical experts or by people specifically trained for that task.
Moreover, the visual interpretation of the images can be subjective, and there may be relevant differences between the diagnoses of different analysts.
All this motivates the research on automatic diagnostic methods~\cite{Pead_SO_2019,Ting_PRER_2019}.

Despite some methods have approached the grading of AMD using retinography images~\cite{Burlina_JAMAO_2018,Tan_FGCS_2018}, the objective of most methods is the diagnosis of referable AMD~\cite{Pead_SO_2019,Ting_PRER_2019}, as it is the most relevant issue in screening.
Referable AMD comprises intermediate and late AMD, but not early AMD. The diagnosis of referable AMD is the focus of the work herein described.
The common approach in the state of the art for AMD diagnosis follows a machine learning-based binary AMD/non-AMD classification.
While early works were based on classical methods with ad hoc features~\cite{Mookiah_KBS_2015,Pead_SO_2019}, most recent works apply Convolutional Neural Networks (CNNs)~\cite{Burlina_JAMAO_2017,Gonzalez-Gonzalo_AO_2020,Li_TMI_2020,Tan_FGCS_2018,Ting_JAMA_2017}, as in this work.
These neural network approaches have explored the use of ad hoc CNN architectures~\cite{Tan_FGCS_2018}, ensembles of these networks~\cite{Ting_JAMA_2017,Gonzalez-Gonzalo_AO_2020}, or standard CNN classification architectures~\cite{Burlina_JAMAO_2017,Li_TMI_2020}.
Moreover, while ImageNet pretraining is common when using standard CNNs~\cite{Burlina_JAMAO_2017,Ting_PRER_2019}, other kinds of self-supervised pretraining were also successfully applied~\cite{Li_TMI_2020}.
All of these are common approaches in the diagnosis of ocular diseases~\cite{Gonzalez-Gonzalo_AO_2020,Ting_JAMA_2017,Ting_PRER_2019}.

In this work, we approach the AMD detection using an ImageNet pretrained CNN~\cite{Simonyan_ICLR_2015}.
In contrast to previous approaches, we propose the use of image-level lesion labels along with diagnosis labels to train a CNN that simultaneously identifies AMD and its associated retinal lesions.
This simultaneous task was not yet addressed in this domain.
Some previous works have approached AMD diagnosis through the identification of certain lesions, such as drusen~\cite{Zheng_ISBI_2013} or geographic atrophy~\cite{Liefers_O_2020}.
In these works, the lesions are first segmented in the image, and then, the AMD diagnosis is derived from that segmentation maps.
Differently, we aim at detecting the image-level presence of a wide variety of lesions associated with AMD, along with the overall AMD diagnosis, in a multi-task learning setting.

The proposed setting has several advantages.
First, it allows to incorporate useful information that conveniently complements the diagnosis, increasing the feedback received by the models during the training.
This multi-task feedback of highly related tasks can help the models to generalize better and to improve the diagnostic performance~\cite{Caruana_ML_1997}.
Second, the lesion presence information provided by the models is of clinical interest, as it can be indicative of the stage of AMD or the presence of other diseases distinct of AMD.
Moreover, due to the direct link between the lesions and the diagnosis, the lesion information can also help to better understand the decisions made by the automatic system, improving its explainability.
Since the model outputs both the diagnosis and the lesion detection, it is easier for clinicians to understand the final diagnosis.
Last, the benefits of this approach are achieved without much extra effort from clinicians for the creation of the training datasets, since the image-level lesion presence identification (lesion labels) is implicit in the diagnosis assessment.
Furthermore, these type of image-level annotations are commonly available in medical records.

To evaluate the proposed approach, we trained a CNN using retinographies and image-level labels from a public dataset, and evaluated its performance in other two public datasets without using any lesion information nor further refinement of the CNN.
The provided ablation study and comparison experiments clearly demonstrate that the proposed approach achieves state-of-the-art performance, and surpasses the performance of the traditional AMD diagnosis approach, while achieving satisfactory results in the identification of most lesions.
Also, a more detailed analysis of the networks outcomes proves that the proposed approach contributes to the explainability of the models.

\section{Materials and Methods}

\begin{figure}[tbph]
    \centering
    \includegraphics[width=\textwidth]{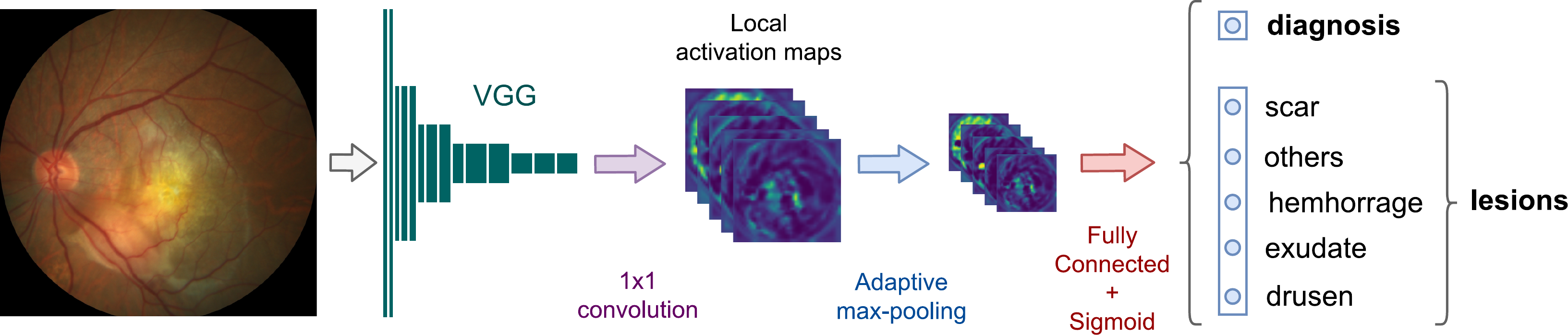}
    \caption{Proposed approach for the simultaneous detection of AMD and the identification of its associated retinal lesions.
    }
    \label{fig:VGG13}
\end{figure}

This work is focused on the simultaneous detection of AMD and the identification of its associated retinal lesions from retinography images. 
To perform this joint task, we train the single CNN depicted in Figure~\ref{fig:VGG13} end-to-end, from raw RGB retinographies.
This joint identification involves predicting the presence/absence of AMD along with the presence/absence of $N$ lesions for each image (AMD + lesions, \pl{A}{L}).
An image can be associated with any number of lesions between 0 and $N$, and the presence of lesions does not necessarily imply the presence of AMD, as they may belong to a different pathology.
Thus, the network outputs correspond to $N+1$ independent image-level detectors.

\subsection{Prediction Loss}

To train the networks using the proposed approach, we use a loss function that combines the diagnosis error and the lesion detection error by means of a weighted sum.
Formally, the loss is defined as
\begin{equation}
     \mathcal{L}_{\textit{\scriptsize total}} ~=~ \mathcal{L}_{\textit{\scriptsize diagnosis}}  ~+~  \alpha~\mathcal{L}_{\textit{\scriptsize lesions}} ~~,
\end{equation}
where $\alpha$ is the weight controlling the relative importance of the diagnosis and the lesion losses, which are defined as
\begin{eqnarray}
    \mathcal{L}_{\textit{\scriptsize diagnosis}} &~=~& \mathcal{L}\left(\textbf{f}(\textbf{r})_1, \textbf{d}\right)~~, \\
    \mathcal{L}_{\textit{\scriptsize lesions}} &~=~& \frac{1}{N}\sum^{N}_{i=1}\mathcal{L}\left(\textbf{f}(\textbf{r})_{i+1}, \textbf{l}_i\right)~~,
\end{eqnarray}
where $\textbf{f}(\textbf{r})$ denotes the predicted network output for retinography $\textbf{r}$,  $\textbf{d}$ the target AMD diagnosis, $\textbf{l}_i$ the target for the identification of lesion $i$,  $N$ the number of lesions, and $\mathcal{L}$ a base binary classification loss.

In the proposed approach the base loss $\mathcal{L}$ is Binary Cross-Entropy, and the number of lesions is $N=5$.
The considered lesions are \textit{drusen}, \textit{exudate}, \textit{hemorrhage}, \textit{scar} and \textit{others} (unknown).
Also, to emphasize the importance of AMD diagnosis, we assign twice as much weight to it in the loss by setting $\alpha=0.5$.

\subsection{Network Architecture}

The proposed CNN architecture is based on the VGG-13 backbone~\cite{Simonyan_ICLR_2015}, as depicted in Figure~\ref{fig:VGG13}.
The convolutional trunk, denoted as VGG in Figure~\ref{fig:VGG13}, corresponds to the convolutional blocks in VGG-13 before the last pooling layer.
This allows to reuse the pre-trained weights in ImageNet classification as initialization.
Moreover, this convolutional block can be directly applied to images of arbitrary size.
The rest of the original VGG-13 network is replaced by a custom pooling and classification head.
First, we add an $N$ channel $1 \times 1$ convolutional layer with ReLU activation to reduce the $512$ channels of the previous layers into $N$ channels---one channel per considered lesion.
This is intended to provide local activation maps of approximately $1/16$ of the original image size on each dimension.
Then, these $N$ maps are reduced to a fixed size of $31 \times 31$ using an adaptive max pooling.
Finally, the output layer consists on a fully-connected layer of $N+1$ sigmoid units  corresponding to each lesion identification and the AMD diagnosis.

\subsection{Data}

For the experiments conducted in this work, we employed the publicly available iChallenge-AMD~\cite{Fu_IEEEDataport_2020}, ARIA~\cite{Farnell_JFI_2008} and STARE~\cite{Hoover_TMI_2000} datasets. 
iChallenge-AMD is used both to train and validate the models in the lesion identification and AMD detection tasks. ARIA and STARE are only used to validate the models in AMD detection.


\paragraph{\textbf{iChallenge-AMD.}}
The iChallenge-AMD dataset~\cite{Fu_IEEEDataport_2020} is composed of 400 images, of which 89 are from patients with AMD.
The size of some images is $2124 \times 2056$ pixels, and $1444 \times 1444$ pixels for others.
All the images are manually labeled as AMD or non-AMD.
The reference standard for AMD presence is based on the retinographies themselves and other complementary information, such as optical coherence tomography (OCT) and visual field.
This information, however, was not released with the dataset and it is not available.
Along with the AMD labels, 118 of the images include pixel-level lesion annotations.
Each lesion is labeled with one of the following classes: drusen (61 images), exudate (38), hemorrhage (19), scar (13) and others (17). The presence of a lesion does not necessarily indicate the presence of AMD, and vice versa.
Although the dataset description does not indicate it, we have found that at least 125 of its images belong to the same eye as another image in the dataset.
This issue is taken into account when partitioning the data into training and test to avoid having images of the same patient in both sets.


\paragraph{\textbf{ARIA.}}
The public ARIA dataset~\cite{Farnell_JFI_2008} is composed of 143 color fundus images from patients with AMD (23), diabetic retinopathy (59) and without any disease (61).
Image sizes are $768 \times 576$ pixels.
To validate our models, we use the 23 images labeled as AMD and the 61 images from healthy patients.


\paragraph{\textbf{STARE.}}
The publicly available STARE dataset~\cite{Hoover_TMI_2000} contains 397 retinographies, of size  $700 \times 605$, from both healthy and pathological patients, with an associated comment indicating its diagnosis.
To validate our approach, we use the 46 images labeled as AMD and the 36 images labeled as ``Normal''.\\~\\

\begin{figure}[tbph]
    \centering
    \includegraphics[width=1.0\textwidth]{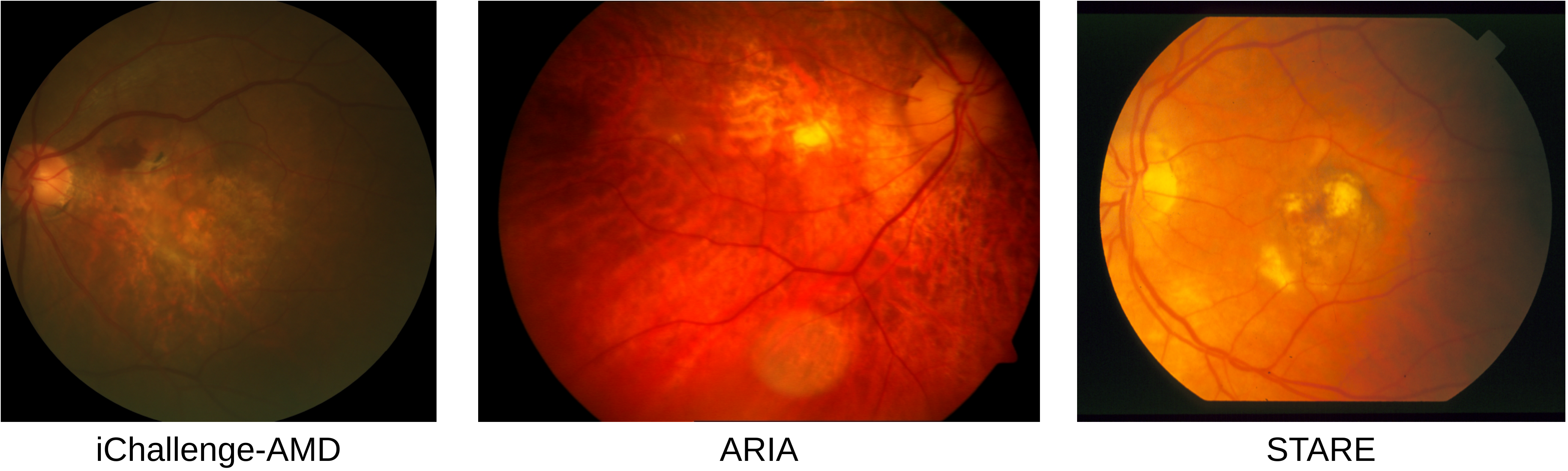}
    \caption{Representative examples of retinography images from the iChallenge-AMD, ARIA and STARE datasets.}
    \label{fig:images_examples}
\end{figure}

In Figure~\ref{fig:images_examples}, representative examples of retinography images from the iChallenge-AMD, ARIA and STARE datasets are provided.

\subsection{Training Details}

For both training and inference, we rescale the iChallenge-AMD dataset images to a fixed width of 720 pixels.
The ARIA and STARE images are preserved at their original resolution, since their width is similar.
To mitigate data scarcity and avoid overfitting, we use online data augmentation.
On each epoch, random transformations are applied to the input images.
These transformations consist of combinations of random intensity and color variations, slight affine transformations (scaling, shearing and rotation) and horizontal and vertical flipping.

To train the different models, we use the Adam optimization algorithm~\cite{Kingma_ICLR_2015}. The parameters of the algorithm were empirically set as follows.
Initial learning rate: $\alpha = 1 \times 10^{-6}$; decay rates: $\beta_1 = 0.9$, $\beta_2 = 0.999$.
The learning rate remains constant during the training, which has a fixed duration of 200 epochs.

The convolutional layers of the networks are initialized to the parameter values of a VGG13 model pre-trained in ImageNet, while the added layers are initialized using the He et al.~\cite{He_ICCV_2015} method with uniform distribution.

\section{Results and Discussion}

To evaluate the impact of including lesion identification feedback, we compare the performance of the models trained with the proposed approach ({\pl{A}{L}}) with the performance of the models trained with the traditional approach, which only involves predicting the presence of AMD (AMD-only, {\mi{A}{O}}).
For this end, we trained the same models using both approaches in the iChallenge-AMD dataset. 
In addition, to take into account the stochasticity of the networks training---magnified by the scarcity of annotated data---we performed 5 repetitions of 2-fold cross-validation with randomly created folds for each considered alternative.
All these models were then also evaluated in the STARE and ARIA datasets.

\begin{figure}[tbph]
    \centering
    \includegraphics[width=\textwidth]{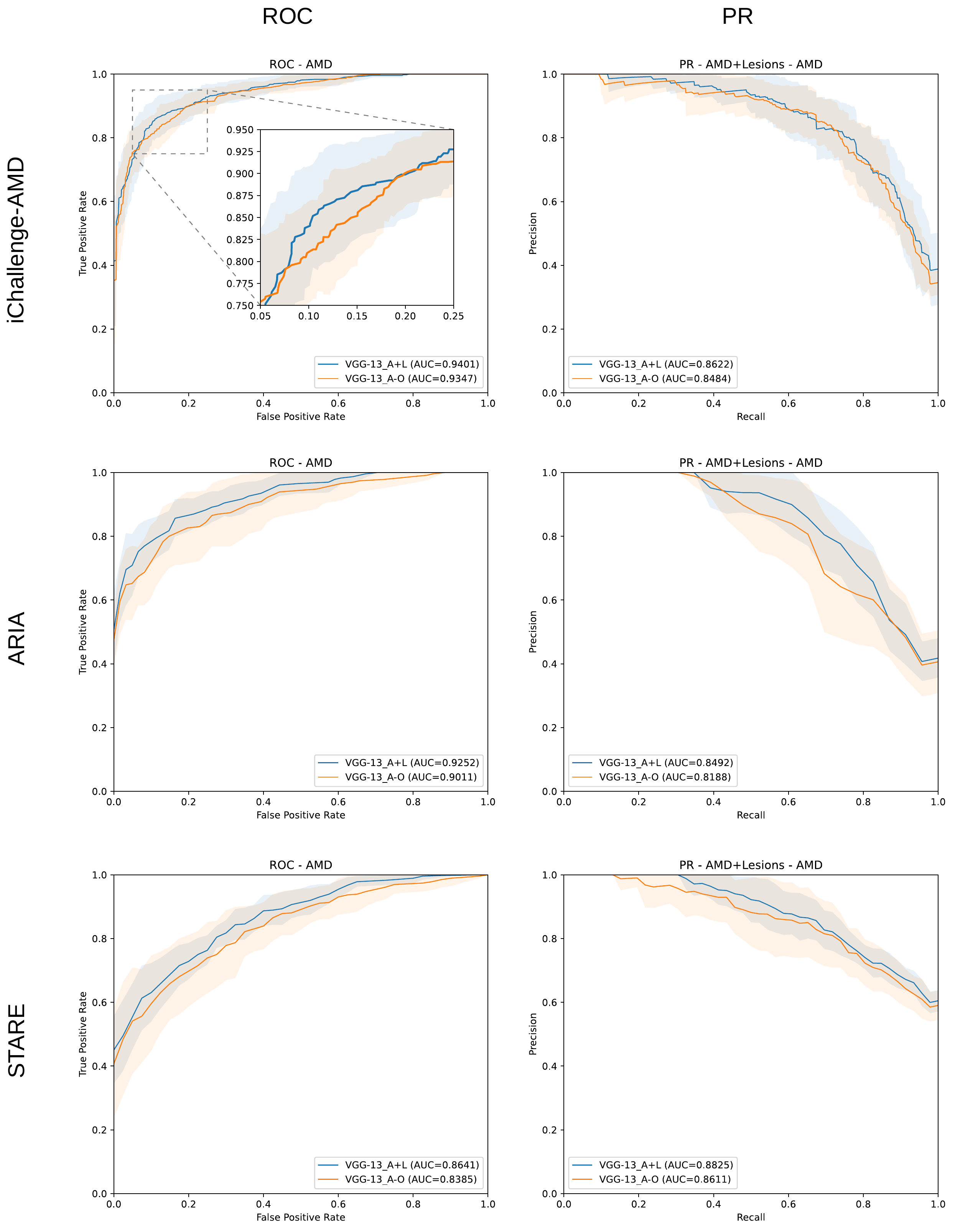}
    \caption{Mean ROC and PR curves for AMD detection for the networks trained using \pl{A}{L} and \mi{A}{O} in the iChallenge-AMD, ARIA and STARE datasets.}
    \label{fig:AMD_curves}
\end{figure}

\begin{figure}[tbph]
    \centering
    \includegraphics[width=\textwidth]{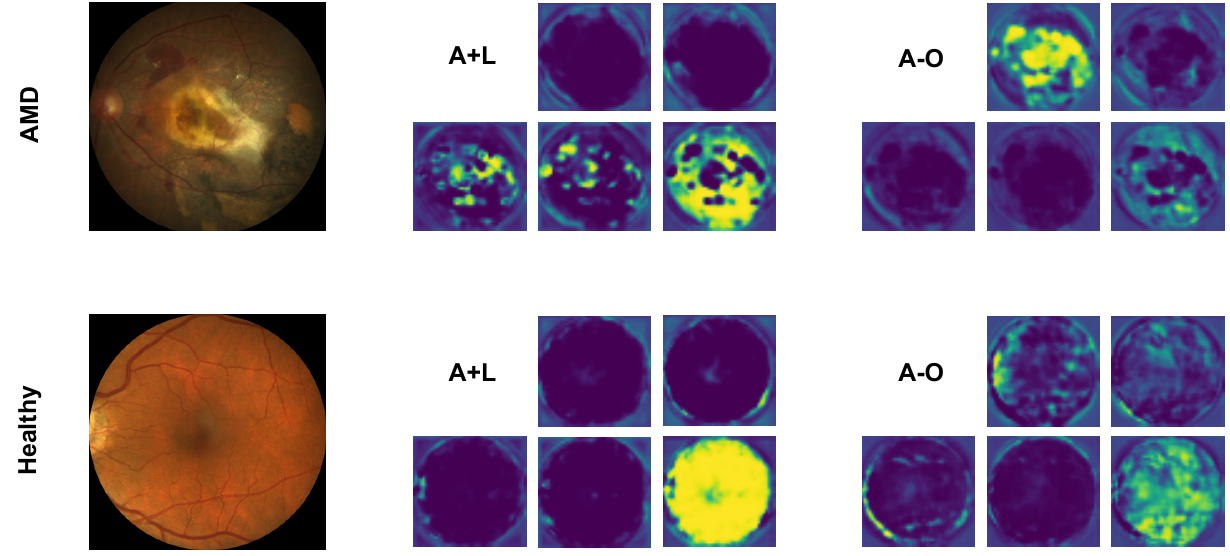}
    \caption{iChallenge-AMD retinographies and its corresponding activation maps after the final $1 \times 1$ convolution for the {\pl{A}{L}} and {\mi{A}{O}} approaches.}
    \label{fig:examples_AL_AO}
\end{figure}

The quantitative evaluation of the models was performed using Precision-Recall (PR) and Receiver Operating Characteristic (ROC) analyses. 
Since a total of 10 repetitions is made for each model, we computed the mean curves by merging each operating point, and computed Area Under the Curve (AUC) values for each curve.
Table \ref{tab:classification_results} reports the AUC values of the mean ROC and PR curves for the \pl{A}{L} and the \mi{A}{O} approaches in the AMD detection task.
In the same table, we include the state-of-the-art works that report their results on the iChallenge-AMD, STARE and ARIA datasets. 
\begin{table}[tbp]
    \centering
    \caption{AMD detection results. All values indicate percentages.}
    \label{tab:classification_results}
    \begin{tabular}{@{\extracolsep{8pt}}llllll}

    \toprule
    Dataset & Metric & 
    \pl{A}{L} & \mi{A}{O} & Li et al.~\cite{Li_TMI_2020} & Mookiah et al.~\cite{Mookiah_KBS_2015} \\
    \midrule
    \multirow{2}{*}{iChallenge-AMD} & AUC-ROC & 94.01 & 93.47 & 77.19 & - \\
    & AUC-PR & 86.22 & 84.84 & - & -  \\
    \midrule
   \multirow{2}{*}{STARE} & AUC-ROC & 86.41 & 83.85 & - & 100 \\
    & AUC-PR & 88.25 & 86.11 & - & - \\
    \midrule
   \multirow{2}{*}{ARIA} & AUC-ROC & 92.52 & 90.11 & - & 85\\
    & AUC-PR & 84.92 & 81.88 & - & - \\

    \bottomrule
    \end{tabular}%
\end{table}
On the other hand, Figure~\ref{fig:AMD_curves} depicts the mean ROC and PR curves of the same models in the iChallenge-AMD, ARIA and STARE datasets, with their corresponding AUC. 
Also, two examples of correctly classified images of the iChallenge-AMD, along with their corresponding lesion activation maps using the \pl{A}{L} and {\mi{A}{O}} approaches, are shown in Figure~\ref{fig:examples_AL_AO}.
Lastly, Table \ref{tab:classification_results_lesions} shows, for the \pl{A}{L} alternative, the AUC values of the mean ROC curves for lesion identification.
\begin{table}[tbp]
    \centering
    \caption{Lesions detection results for the \pl{A}{L} approach on iChallenge-AMD.}
    \label{tab:classification_results_lesions}
    \begin{tabular}{@{\extracolsep{8pt}}llllll}
    \toprule
    \multicolumn{1}{l}{} & Drusen & Exudate & Hemorrhage & Scar & Others \\ \midrule
    \multicolumn{1}{l}{AUC-ROC (\%)} & 71.43 & 85.87 &  85.67 & 89.29 &  41.61  \\ \bottomrule
    \end{tabular}%
\end{table}

As it can be observed in Table~\ref{tab:classification_results} and Figure~\ref{fig:AMD_curves}, the best results are achieved by the \pl{A}{L} models.
The difference between these models increases for STARE and ARIA datasets, used only for validation.
This demonstrates that incorporating lesion identification feedback helps the models to generalize better.
In addition, both the \pl{A}{L} and the \mi{A}{O} approaches clearly outperform the works in the state of the art using the iChallenge-AMD and ARIA datasets.
In STARE, however, this is not the case, and the work by Mookiah et al.~\cite{Mookiah_KBS_2015} is the one that yields the best results.
In this regard, however, it is worth noting that the results shown in~\cite{Mookiah_KBS_2015} correspond to a model directly trained in STARE, while ours correspond to a model trained on a different dataset (iChallenge-AMD), without further refinement in STARE.

As observed in Figure~\ref{fig:examples_AL_AO}, the \pl{A}{L} CNN local activation maps clearly differ between pathological and healthy images.
In the latter case, the overall highest values along the whole image belong to a single map.
Differently, for pathological images, these maximum values are distributed among several maps. Furthermore, these values are commonly placed in areas where the lesions occur. This is the same for all the models of the \pl{A}{L} approach.
Thus, in a sense, the activations of the last map (bottom right corner) represent the healthy areas of the image, while the activations of the other maps indicate the presence of a lesion. This information is of clinical interest, as it indicates the areas in which the model estimates that a lesion is placed. Also, it helps to better understand the decisions made by the model, increasing its explainability and facilitating its evaluation.
Also, as can be seen in Figure~\ref{fig:examples_AL_AO}, the {\pl{A}{L}} approach gives rise to more informative activation maps, as it approximately differentiates the lesions from each other and provides more precise activation regions.

Related to this, the results in Table~\ref{tab:classification_results_lesions} show that the \pl{A}{L} models are able to identify the exudate, hemorrhage and scar lesions with satisfactory results.
We observe, however, a remarkable lower performance for the `drusen' and `other' lesions. The low performance on drusen is explained by their usual subtle appearance, as well as their similarity with exudates.
In the case of the other lesion detector, the achieved performance corresponds to a system with random behavior and unable to extrapolate.
This is due to the great variety of lesions included in this class, some of which are subtle and similar to the other labeled lesions, along with the poor representation of such complexity in the training samples.

In general, the above results lead to two main conclusions.
First, that adding lesion identification feedback during the training of the networks clearly improves the diagnosis. 
And second, that this task also allows the models to provide meaningful probability vectors of the lesions present in an image, being particularly accurate for exudates, hemorrhages and scars.
This information, together with the network activation maps, conveniently complements the diagnosis, and can help to better understand the decisions made by the models.

\section{Conclusions}

In this work, we have proposed a novel approach for improving the AMD diagnosis by the simultaneous identification of its associated retinal lesions. To this end, we trained a CNN using the retinography images and the image-level annotations (both for lesions and diagnosis) from the publicly available iChallenge-AMD dataset.
The experiments performed in this and other two public datasets (STARE and ARIA) demonstrate that the proposed approach surpasses the traditional approach in AMD detection, while achieving satisfactory results regarding the identification of most lesions.
The information resulting from this latter task (the vector of lesion probabilities and the network activation maps) conveniently complements the diagnosis, and can be useful to better understand the decisions made by the model.
Moreover, collecting the data needed for this task does not imply much extra effort from clinicians, since the identification of lesions is integrated into the diagnostic process.
Apart from this, there are currently many datasets that, while not providing this type of annotations, do provide pixel-level lesion annotations for a subset of its data, from which the image-level lesion labels can be easily obtained.
In light of the results herein presented, we think that the proposed methodology could be successfully applied in most of these scenarios.

Notwithstanding, our approach also presents some potential issues for further improvement.
The most important one is related to lesion identification. Based on the provided results, the capability of our model to identify the selected lesions could be substantially improved, particularly for the most difficult ones (drusen and others).
In this regard, one of the aspects that could help to improve the performance is the addition of more data, since our approach is trained with very few examples.
On the other hand, the activation maps that are provided by our approach, although useful, are not linked to the lesions, which complicates its analysis.
For this to be the case, new architectural designs need to be explored. Both issues represent interesting fields for further research.

\subsubsection{Acknowledgments.}
{
This work was funded by Instituto de Salud Carlos III, Government of Spain, and the European Regional Development Fund (ERDF) of the European Union (EU) through the \mbox{DTS18/00136} research project;
Ministerio de Ciencia e Innovación, Government of Spain, through \mbox{RTI2018-095894-B-I00} and \mbox{PID2019-108435RB-I00} research projects;
Axencia Galega de Innovación (GAIN), Xunta de Galicia, ref. \mbox{IN845D 2020/38};
Xunta de Galicia and the European Social Fund (ESF) of the EU through the predoctoral grant contracts ref.
\mbox{ED481A-2017/328}
and ref. \mbox{ED481A 2021/140};
Consellería  de  Cultura,  Educación e Universidade, Xunta de Galicia, through Grupos de Referencia Competitiva, grant ref. \mbox{ED431C 2020/24};
CITIC, Centro de Investigación de Galicia ref. \mbox{ED431G 2019/01}, is funded by Consellería de Educación, Universidade e Formación Profesional, Xunta de Galicia, through the ERDF (80\%) and Secretaría Xeral de Universidades (20\%).
}


\clearpage

\bibliographystyle{splncs04}
\bibliography{references}

\end{document}